\documentclass[galaxies,article,accept,moreauthors,pdftex,12pt,a4paper]{mdpi}

\usepackage{amsmath,amssymb}
\usepackage{graphics}
\usepackage{soul}

\lastpage{519}
\doinum{10.3390/galaxies2040496}
\pubvolume{2 (4)}
\pubyear{2014}
\history{Received: 13 September 2014; in revised form: 15 October 2014 / Accepted: 20 October 2014 / \\Published: 28 October 2014}

\Title{Bianchi Type I Cosmological Models in Eddington-inspired Born--Infeld Gravity}

\Author{Tiberiu Harko $^{1}$, Francisco S.N.~Lobo $^{2,}$* and Man Kwong Mak $^{3}$}

\address{%
$^{1}$ Department of Mathematics, University College London, Gower Street, London WC1E 6BT, UK; E-Mail: t.harko@ucl.ac.uk \\
$^{2}$ Centro de Astronomia e Astrof\'{\i}sica da Universidade de Lisboa, Campo Grande, \mbox{Ed. C8 1749-016 Lisboa,} Portugal\\
$^{3}$ Department of Computing and Information Management, Hong Kong Institute of Vocational Education, Chai Wan, Hong Kong, China; E-Mail: mkmak@vtc.edu.hk

}

\corres{E-Mail: flobo@cii.fc.ul.pt; \linebreak {Tel.: +351 217500986}.

\vspace{12pt}
External Editor: {{Antonaldo Diaferio }}

}

\abstract{ We consider the dynamics of a barotropic cosmological fluid in an anisotropic, Bianchi type I space-time in Eddington-inspired Born--Infeld (EiBI) gravity. By assuming isotropic pressure distribution, we obtain the general solution of the field equations in an exact parametric form. The behavior of the geometric and thermodynamic parameters of the Bianchi type I Universe is studied, by using both analytical and numerical methods, for some classes of high density matter, described by the stiff causal, radiation, and pressureless fluid equations of state. In all cases the study of the models with different equations of state can be reduced to the integration of a highly nonlinear second order ordinary differential equation for the energy density. The time evolution of the anisotropic Bianchi type I Universe strongly depends on the initial values of the energy density and of the Hubble function. An important observational parameter, the mean anisotropy parameter, is also studied in detail, and we show that for the dust filled Universe the cosmological evolution always ends into isotropic phase, while for high density matter filled universes the isotropization of Bianchi type I universes is essentially determined by the initial conditions of the energy density.\\\\}

\keyword{Bianchi type I models; Eddington-inspired Born--Infeld gravity; modified gravity; cosmology}

\def\be{\begin{equation}}
\def\ee{\end{equation}}
\def\bea{\begin{eqnarray}}
\def\eea{\end{eqnarray}}

\def\bl{\begin{align}}
\def\el{\end{align}}

\begin{document}

\vspace{-12pt}
\section{Introduction}

The cosmological principle, representing the cornerstone of present day cosmology, assumes that at large scales the Universe is statistically isotropic and homogeneous. The cosmological principle can be tested observationally, and one possibility is through the study of the anisotropies in the Cosmic Microwave Background (CMB) radiation. It is interesting to note that the recently released Planck satellite data \cite{Plancka,Planckb,Planckc} show a significant deviation from large scale isotropy (around 3$\sigma $). The Wilkinson Microwave Anisotropy Probe (WMAP) has also previously found evidence for the deviations from isotropy \cite{WMAP}. Other recently found large-scale anomalies in the maps of temperature anisotropies in the cosmic microwave background include alignments of the largest modes of CMB anisotropy with each other and with geometry, and the direction of motion of the Solar System, and the unusually low power at these largest scales \cite{LS,Campanelli:2006vb,Campanelli:2007qn}. Therefore, by taking into account the present day observational results, \mbox{the possibility} that the large scale structure of the Universe is not statistically perfectly isotropic cannot be ruled out {\it a priori}.

\scalebox{.98}{On the other hand, the firm observational confirmation of the late-time acceleration of the Universe \cite{Riess}} has posed a major challenge to the theoretical foundations of cosmology, namely, General Relativity (GR). The explanation of the de Sitter type acceleration requires the introduction of either a cosmological constant, or of a mysterious dark energy \cite{Am, Mi}, filling the Universe and dominating its expansionary evolution. Both of these components lack a firm theoretical basis and convincing observational evidence for their existence. Therefore one possibility to explain the accelerated expansion of the Universe is to assume that it is due to purely geometric effects, and that a more general theory, having general relativity as a limiting case, describes the large scale dynamics and evolution of the Universe. Several, essentially geometric, extensions of the standard general relativity model have been considered and investigated in detail as alternatives to dark energy and the cosmological constant. Some of the recently considered, and extensively investigated, geometric modifications of general relativity that can explain the late de Sitter type expansionary phase in the dynamical evolution of the Universe include: the $f(R)$ type generalized gravity models \cite{reva,revb,revc,revd,reve}, where $R$ is the Ricci scalar; the $f\left(R,L_m\right)$ models with curvature-matter coupling \cite{coupla,coup1b,coup1c}, where $L_m$ is the matter Lagrangian; the $f(R,T)$ models \cite{fRT}, where $T$ is the trace of the energy-momentum tensor; the Weyl--Cartan--Weitzenb\"{o}ck (WCW) gravity \cite{WCW}; hybrid metric-Palatini $f(X)$ type models \cite{fXa,fXb,fXc,fXd,fXe}; or $f\left(R,T,R_{\mu \nu}T^{\mu \nu }\right)$ gravity \cite{fRTT}, where $R_{\mu \nu}$ is the Ricci tensor and $T_{\mu \nu }$ is the matter energy-momentum tensor. For a recent review of the generalized curvature-matter couplings, in particular, $f\left(R,L_m\right)$ and $f(R,T)$ modified theories of gravity, we refer the reader to \cite{galrev}.

In standard GR, the coupling between matter and gravity is given by a proportionality relation between the energy-momentum tensor and the geometry, so that $G_{\mu \nu}\sim T_{\mu \nu}$, where $G_{\mu \nu}$ is the Einstein tensor. Although both the energy-momentum tensor and the Einstein tensor are divergenceless, there is no obvious reason why the matter--gravity coupling should be linear. On the other hand, modified theories of gravity usually affect the vacuum dynamics, yet keep the matter--gravity coupling linear. Based on the early work of Eddington \cite{4}, Born and Infeld \cite{5}, and Deser and Gibbons \cite{Deser}, the so-called Eddington-inspired Born--Infeld (EiBI) theory has been recently proposed in \cite{1} to provide such coupling modifications, by resurrecting Eddington's old proposal for the gravitational action in the presence of a cosmological constant, and extending it to include matter fields. In the presence of sources the Poisson equation is modified, and charged black holes show great similarities with those arising in Born--Infeld electrodynamics coupled to gravity. Interesting cosmological consequences also appear when one considers homogeneous and isotropic space-times. In this case it turns out that there is a minimum length (and a maximum density) at early times, showing that the EiBI theory can provide an alternative theory of the Big Bang, and with a non-singular description of the Universe. {The theory also introduces a coupling parameter $\kappa $ (the Eddington parameter), a constant with inverse dimensions to that of a cosmological constant. When the parameter $\kappa \ll |g|/R$, where $|g|$ is the determinant of the metric tensor and $R$ is the Ricci scalar, the EiBI action reduces to the Einstein--Hilbert action with cosmological constant $\Lambda $. When $\kappa \gg |g|/R$, the Eddington action will be obtained approximately. Therefore, the EiBI parameter $\kappa $ interpolates between two different gravity theories.}

The astrophysical and cosmological consequences of the EiBI theory have been extensively investigated recently. For a positive coupling parameter, the field equations have a dramatic impact on the collapse of dust, and do not lead to singularities \cite{Sot1}. The theory supports stable, compact pressureless stars made of perfect fluid, which provide interesting models of self-gravitating dark matter. The existence of relativistic stars imposes a strong, near optimal constraint on the coupling parameter, which can even be improved by observations of the moment of inertia of the double pulsar \cite{Sot1}. In \cite{2}, it was proven that the EiBI theory coupled to a perfect fluid reduces to General Relativity coupled to a nonlinearly modified perfect fluid, leading to an ambiguity between modified coupling and modified equation of state. The observational consequences of this degeneracy were discussed, and it was shown that such a completion of General Relativity is viable from both an experimental and theoretical point of view through the energy conditions, consistency, and singularity-avoidance perspectives. \mbox{In \cite{3}} it was shown that the EiBI theory is reminiscent of Palatini $f(R)$ gravity, and that it shares the same pathologies, such as curvature singularities at the surface of polytropic stars, and unacceptable Newtonian limit. The dark matter properties in the EiBI theory were considered in \cite{Harko}. The properties of the stellar type objects as well some spherically symmetric models were considered \mbox{in \cite{prop1a,prop1b,prop1c,prop1d,prop1e,prop1f}}. Wormhole solutions within the framework of the theory were obtained in \cite{worm}. Other theoretical and astrophysical implications of the EiBI theory were investigated recently in \cite{ex1,ex2,ex3,ex4,ex5,ex6,ex7,ex8,ex9,ex10,ex11,ex12,ex13,ex14,ex15}.

\scalebox{.975}{The cosmological implications of the EiBI theory have also been investigated. The tensor perturbations} of a homogeneous and isotropic space-time in the Eddington regime, where modifications to Einstein gravity are strong, were studied in \cite{cosm1}. The tensor mode is linearly unstable deep in the Eddington regime, and, even though the background evolution is resolutely non-singular, the overall cosmological evolution is still singular, once one considers tensor perturbations. The evolution of \mbox{a Universe} permeated by a perfect fluid with an arbitrary equation of state parameter $w$ was analyzed in \cite{cosm2}. \mbox{A bounce} may occur for $\kappa >0$, if $w$ is time-dependent, and this model is free from tensor singularities. Hence EiBI cosmologies may provide a viable alternative to the inflationary paradigm, as a solution to the fundamental problems of the standard cosmological model. The evolution of the Universe filled with barotropic perfect fluid in the EiBI theory was considered in \cite{cosm3}, for both isotropic and anisotropic Universes. At the early stages, when the energy density is high, the evolution is considerably modified as compared with that in general relativity. For the equation-of-state parameter $w>0$, there is no initial singularity and for pressureless dust ($w=0$), the initial state approaches a de Sitter type evolution. The anisotropy is mild, and does not develop curvature singularities in space-time. The dynamics of homogeneous and isotropic Universes was also further explored \cite{cosm4}. For $\kappa > 0$ there is a singularity-avoiding behavior in the case of a perfect fluid with equation of state parameter $w > 0$. The range $-1/3 < w < 0$ leads to Universes that experience unbounded expansion rate, whilst still at a finite density. In the case $\kappa < 0$ the addition of spatial curvature leads to the possibility of oscillation between two finite densities. Domination by a scalar field with an exponential potential also leads to singularity-avoiding behavior when $\kappa > 0$. The behavior of a homogeneous and isotropic Universe filled with phantom energy in addition to the dark and baryonic matter was analyzed in \cite{cosm5,Bouhmadi-Lopez:2014jfa,Bouhmadi-Lopez:2014tna}. Unlike the Big Bang singularity that can be avoided through a bounce or a loitering effect on the physical metric, the Big Rip singularity is unavoidable in the EiBI phantom model, even though it can be postponed towards a slightly further future cosmic time. {The evolution of \mbox{a spatially-flat}, homogeneous anisotropic Kasner universe filled with a scalar field, whose potential has various forms, was studied in Eddington-inspired Born--Infeld gravity in \cite{new1}. By imposing a maximal pressure condition, an exact solution for each scalar field potential, describing the initial state of the universe, was found. The initial state is regular if the scalar potential increases no faster than the quadratic power for large field values. Contrary to the case of general relativity, the anisotropy does not generate any defects in \mbox{the early universe.}}

It is the goal of this paper to present a systematic investigation of the simplest anisotropic cosmological model, described by a Bianchi type I geometry, in the framework of EiBI gravity.
We refer the reader to \cite{Rodrigues:2008kv} for related work. After writing down the basic cosmological evolution equations in the anisotropic geometry, we consider three distinct models, corresponding to three different choices of the equation of state of the matter. More exactly, we investigate the cosmological dynamics for universes filled with a stiff fluid, a radiation fluid and dust, respectively. In all these cases we obtain the basic evolution equation of the model, which in general is described by a highly nonlinear second order ordinary differential equation. The time dynamics is studied by using both approximate analytical and numerical methods. The evolution of the anisotropic Bianchi type I Universe strongly depends on the initial values of the density and of the Hubble function. An important observational parameter, the mean anisotropy parameter, is studied in detail, and we show that for the dust filled Universe the cosmological evolution always ends in isotropic phase, while for high density matter filled universes the isotropization is essentially determined by the initial conditions. {An alternative possibility of studying the isotropization of a Bianchi type I geometry is via the shear scalar $\sigma ^2=\sigma _{ij}\sigma ^{ij}$, where $\sigma _{ij}$ is the shear tensor. The shear tensor $\sigma _{ij}$ shows the tendency of evolution into an ellipsoidal shape of an initially spherically symmetric region. Hence, the shear scalar $\sigma ^2$ gives the distortion rate of a large scale cosmological structure.}

{The Bianchi type I anisotropic geometries represent the simplest extension of the standard Friedmann--Robertson--Walker (FRW) line element, to which they reduce in the limit of equal scale factors. Bianchi type models can be considered as viable alternatives to the standard FRW isotropic geometry, with small deviations from the exact isotropy that could explain the anisotropies and anomalies in the CMB. Motivated by the large scale asymmetry observed in the cosmic microwave background sky, a specific class of anisotropic cosmological models---Bianchi type ${\rm VII}_h$---was considered in \cite{Jaffe}. A comparison with the WMAP first-year data on large angular scales was performed, and evidence of a correlation ruled out as a chance alignment at the $3\sigma $ level was found. However, the recent Planck Collaboration results \cite{Planckb} did show that the Bianchi type ${\rm VII}_h$ cosmological model is not consistent with the observational data obtained by the Planck satellite. On the other hand, one of the large angle anomalies of the CMB, the low quadrupole moment, indicating a great amount of power suppression at large scales, seems to point towards the presence of a Bianchi type I anisotropic geometry. \mbox{The smallness} of the quadrupole component of the CMB temperature distribution implies that if the universe is homogeneous but anisotropic, the deviation from the FRW geometry must be small, and thus such a deviation fits naturally into the framework of the Bianchi type I geometry.}

The present paper is organized as follows. In Section~\ref{sect1}, we briefly review the theoretical foundations of EiBI gravity, and we write down the field equations for an anisotropic Bianchi type I geometry. \mbox{The basic} equations describing the physical and geometric properties of the Bianchi type I models with isotropic pressure distribution are obtained in Section~\ref{sect2}. In Section~\ref{sect3}, explicit Bianchi type I models in EiBI gravity are studied for stiff and radiation fluids, by numerically integrating the cosmological evolution equations. The case of the dust Bianchi type I Universes is considered in Section~\ref{sect4}. We discuss and conclude our results in Section~\ref{sect5}.

\section{Eddington-inspired Born--Infeld Gravity}\label{sect1}

In the present section we write down the action and the field equations of EiBI gravity, and obtain the explicit form of the field equations for a Bianchi type I anisotropic and homogeneous geometry.

\subsection{Gravitational Action and Field Equations}

The starting point of the EiBI theory is the action $S$ given by
\begin{eqnarray}
S=\frac{1}{16\pi }\frac{2}{\kappa }\int d^{4}x\left( \sqrt{\left| g_{\mu
\nu }+\kappa R_{\mu \nu }\right| }-\lambda \sqrt{-g}\right) +S_{M}\left[ g,\Psi _{M}\right] \label{1c}
\end{eqnarray}
where $\lambda \neq 0$ is a dimensionless constant, and $\kappa $ is a parameter with inverse dimension to that of \mbox{the cosmological} constant $\Lambda $. $R_{\mu \nu }$ is the symmetric part of the Ricci tensor, and is constructed solely from the connection $\Gamma _{\beta \gamma }^{\alpha }$.

The following relation is obtained by varying the action (\ref{1c}) with respect to the connection $\Gamma _{\beta \gamma }^{\alpha }$
\begin{equation}
q_{\mu \nu }=g_{\mu \nu }+\kappa R_{\mu \nu } \label{2c}
\end{equation}
Varying the action (\ref{1c}) with respect to the physical metric $g_{\mu \nu }$, with the help of Equation~(\ref{2c}), gives
\begin{equation}
\sqrt{-q} \, q^{\mu \nu }= \sqrt{-g} g^{\mu \nu }- 8\pi \kappa \sqrt{-g}\; T^{\mu \nu } \label{3c}
\end{equation}
where we have introduced the auxiliary metric $q_{\mu \nu }$ related to the connection given by
\begin{equation}
\Gamma _{\beta \gamma }^{\alpha }=\frac{1}{2}q^{\alpha \sigma }\left(
\partial _{\gamma }q_{\sigma \beta }+\partial _{\beta }q_{\sigma \gamma
}-\partial _{\sigma }q_{\beta \gamma }\right) \label{4c}
\end{equation}

The gravitational field equation is given by
\begin{eqnarray}
R_{\nu }^{\mu }&=&8\pi \tau T_{\nu }^{\mu }+\frac{1-\tau }{\kappa }\delta
_{\nu }^{\mu } \label{13c}
\end{eqnarray}
where we have used the relations $R=R_{\mu }^{\mu }$ and $T=T_{\mu }^{\mu }$.

Using $R=8\pi \tau T+\frac{4(1-\tau) }{ \kappa }$, the Einstein tensor $%
G_{\mu}^{\nu }$ for the apparent metric $q_{\mu\nu}$ then follows immediately
\begin{equation}
G_{\nu}^{\mu }=R_{\nu}^{\mu }-\frac{1}{2} \delta_{\nu}^{\mu }R=8\pi S_{\nu}^{\mu } \label{15c}
\end{equation}
where we have denoted the apparent energy momentum tensor $S_{\nu }^{\mu }$
as
\begin{equation}
S_{\nu }^{\mu }=\tau T_{\nu }^{\mu }- \left( \frac{1-\tau}{8\pi \kappa} +
\frac{\tau}{2} T \right) \delta _{\nu }^{\mu } \label{16c}
\end{equation}

Note that the quantity $\tau $ can be obtained from $T_{\nu }^{\mu }$ by
\begin{equation}
\tau =\left| \delta _{\nu }^{\mu }-8\pi \kappa T_{\nu }^{\mu }\right| ^{-%
\frac{1}{2}} \label{18c}
\end{equation}
which can be expressed in terms of physical quantities given by
\begin{equation}
\tau =\left[ \left( 1+8\pi \kappa \rho \right) \left( 1-8\pi \kappa p_1\right) \left( 1-8\pi \kappa p_2\right) \left( 1-8\pi \kappa p_3\right) %
\right] ^{-\frac{1}{2}} \label{19c}
\end{equation}
where the pressures $p_i$ with $i=1,2,3$, are defined along the $x$, $y$ and $z$ directions, respectively.

\subsection{The Gravitational Field Equations for a Bianchi Type I Geometry in EiBI Gravity}

The Bianchi type I metric line elements for the real metric $g_{\mu \nu }$ and for the auxiliary metric $q_{\mu \nu }$ take the following forms respectively
\begin{eqnarray} \label{8c}
g_{\mu \nu }dx^{\mu }dx^{\nu }=-dt^{2}+g_1^2(t)dx^{2}
+g_2^2(t)dy^{2}+g_3^2(t)dz^{2} \\
q_{\mu \nu }dx^{\mu }dx^{\nu }=-dt^{2}+a_1^2(t)dx^{2}
+a_2^2(t)dy^{2}+a_3^2(t)dz^{2} \label{9c}
\end{eqnarray}

The full system of the gravitational field equations for a Bianchi type I space-time in the EiBI gravity model are given by
\begin{eqnarray}
\frac{1}{ \kappa}\left( 1-\frac{A}{B_{1}B_{2}B_{3}}\right) = \frac{\ddot{a}_1 a_2 a_3 + a_1 \ddot{a}_2 a_3 + a_1 a_2 \ddot{a}_3}{a_1 a_2 a_3 } \label{S2tt} \\
\frac{1}{\kappa}\left( 1-\frac{B_1}{AB_{2}B_{3}}\right) = \frac{\dot{a}_1 \dot{a}_2 a_3 + \dot{a}_1 a_2 \dot{a}_3 + \ddot{a}_1 a_2 a_3}{a_1 a_2 a_3} \label{S2rr} \\
\frac{1}{\kappa}\left( 1-\frac{B_2}{AB_{1}B_{3}}\right) = \frac{\dot{a}_1 \dot{a}_2 a_3 + a_1 \dot{a}_2 \dot{a}_3 + a_1 \ddot{a}_2 a_3}{a_1 a_2 a_3 } \\
\frac{1}{ \kappa}\left( 1-\frac{B_3}{AB_{1}B_{2}}\right)= \frac{\dot{a}_1 a_2 \dot{a}_3 + a_1 \dot{a}_2 \dot{a}_3 + a_1 a_2 \ddot{a}_3}{a_1 a_2 a_3 }
\label{Stheta}
\end{eqnarray}

From the field Equation (\ref{3c}), we obtain
\begin{equation}
a_i=g_i \frac{A}{B_i}
\end{equation}
where we have defined the arbitrary functions $A(t)$ and $B_{i}(t)$ ($i=1,2,3$) respectively:
\begin{eqnarray}
A^{2} &=&1+8\pi \kappa \rho \label{AA} \\
B_{i}^{2} &=&1-8\pi \kappa p_{i}, \qquad i=1,2,3
\end{eqnarray}%

\section{Isotropic Pressure Bianchi Type I Universes in EiBI Gravity}\label{sect2}

Of particular importance for the understanding of the global cosmological dynamics are the isotropic pressure Bianchi type I models, for which
\begin{equation}
p_{1}=p_{2}=p_{3}=p
\end{equation}

Therefore
\begin{equation}
B_{1}=B_{2}=B_{3}=B=\sqrt{1-8\pi \kappa p}
\end{equation}

With the help of the scale factors one can define the following new variables \cite{Gron}
\begin{eqnarray}
V &=&\prod _{i=1}^3{a_i}, \qquad H_{i}=\frac{\dot{a}_{i}}{a_{i}}, \qquad i=1,2,3 \notag
\label{var} \\
H &=&\frac{1}{3}\sum_{i=1}^{3}{H_{i}}, \qquad \Delta H_{i}=H-H_{i}, \qquad i=1,2,3
\end{eqnarray}%

In Equation~(\ref{var}), $V$ represents the volume scale factor, $H_{i}$ (with $
i=1,2,3$) are the directional Hubble functions, and $H$ is the mean Hubble function, respectively. By using the definitions of $H$ and $V$ we immediately obtain
\begin{equation}
H=\frac{1}{3}\frac{\dot{V}}{V}
\end{equation}%

In the physical $g$ space the comoving volume element $V^{(g)}=g_1g_2g_3$ is obtained as
\be V^{(g)}=g_1g_2g_3=\frac{B^3}{A^3}V
\ee

As an indicator of the degree of anisotropy of a cosmological model one can take the mean anisotropy parameter $A_{p}$, defined in the $q$ space according to \cite%
{Fabbri}
\begin{equation}
A_{p}=\frac{1}{3}\sum_{i=1}^{3}{\left( \frac{\Delta H_{i}}{H}\right) ^{2}}
\end{equation}%

\mbox{For a cosmological model that is isotropic, $H_{1}=H_{2}=H_{3}=H$ and $A_{p}\equiv 0$, respectively. The anisotropy} parameter is an important indicator of the behavior of anisotropic cosmological models, since in standard four-dimensional general relativity it is finite even for singular states (for example, $A_{p}=2$ for Kasner-type geometries \cite{20}). The time evolution of $A_{p}$ is a good indicator of the dynamics of \linebreak the anisotropy \cite{HMan}.

In the physical $g$ space we define the anisotropy parameter as
\begin{equation}
A_p^{(g)}=\frac{1}{3}\sum_{i=1}^{3}{\left[ \frac{\Delta H_{i}^{(g)}}{H^{(g)}}\right] ^{2}}
\end{equation}

With the use of the variables given by Equation~(\ref{var}), in the $q$-metric of the EiBI theory the gravitational field equations for a Bianchi type I space-time filled with isotropic fluid take the form
\begin{equation}
3\dot{H}+\sum _{i=1}^3{H_{i}^{2}}=\frac{1}{ \kappa }\left( 1-\frac{A}{B^3}\right) \label{f1}
\end{equation}%
\begin{equation}
\frac{1}{V}\frac{d}{dt}\left( VH_{i}\right) =\frac{1}{\kappa }\left( 1-\frac{1}{AB}\right) ,
\qquad i=1,2,3 \label{f2}
\end{equation}

By adding Equation~(\ref{f2}) we find
\begin{equation}
\frac{1}{V}\frac{d}{dt}\left( VH\right) =\dot{H}+3H^{2}=\frac{1}{\kappa }%
\left( 1-\frac{1}{AB}\right) \label{f3}
\end{equation}%

Furthermore, by subtracting Equations~(\ref{f2}) and (\ref{f3}) we obtain
\begin{equation}
\frac{d}{dt}V\left( H_{i}-H\right) =0, \qquad i=1,2,3
\end{equation}%
which provides
\begin{equation}
H_{i}=H+\frac{K_{i}}{V}, \qquad i=1,2,3 \label{41}
\end{equation}%
where $K_{i}$ are arbitrary constants of integration, and
\begin{equation}
a_{i}=a_{i0}V^{1/3}\exp \left[ K_{i}\int {\left(\frac{1}{V}\frac{dt}{dV}\right)dV}\right]
, \qquad i=1,2,3 \label{aa}
\end{equation}%
where $a_{i0}$ are arbitrary constants of integration. From Equation~(%
\ref{41}) it follows that the integration constants $K_{i}$ (with $i=1,2,3$) must satisfy the consistency condition
\begin{equation}
\sum_{i=1}^{3}{K_{i}}=0
\end{equation}%

By substituting the expression of $H=\dot{V}/(3V)$ into Equation~(\ref{f3}), it follows that the function $V$ satisfies the following second order differential equation,
\begin{equation}
\ddot{V}=\frac{3}{\kappa }\left( 1-\frac{1}{AB}\right) V
\label{V}
\end{equation}

By substituting Equation~(\ref{41}) into Equation~(\ref{f1}), and by taking into account the definition of $H$ and Equation~(\ref{f3}), we obtain the relation
\be\label{constr}
-6H^2+\frac{K^2}{V^2}=\frac{1}{\kappa }\left(\frac{3}{AB}-\frac{A}{B^3}-2\right)
\ee where we have denoted $K^2=\sum_{i=1}^3{K_i^2}$. Equation (\ref{V}) can be integrated to give
\begin{equation}\label{time}
t-t_{0}=\int \left\{ C_{0}+\frac{6}{\kappa }\int ^{V}\left[ 1-\frac{1}{A\left(V'\right)B\left(V'\right)}\right] V'dV'\right\} ^{-1/2}dV
\end{equation}
where $t_{0}$ and $C_{0}$ are arbitrary constants of integration. With the use of Equation~(\ref{time}) it follows that Equation~(\ref{constr}) becomes
\be K^2-\frac{2}{3}C_0=\frac{4}{\kappa }\int{\left(1-\frac{1}{AB}\right)VdV}+\frac{1}{\kappa }\left(\frac{3}{AB}-\frac{A}{B^3}-2\right)V^2
\ee thus giving a constraint on the integration constants $K^2$ and $C_0$.

For the directional Hubble parameters in the $g$-space we obtain
\be H_i^{(g)}=\frac{\dot{g}_i}{g_i}=H_i+\frac{\dot{B}}{B}-\frac{\dot{A}}{A}, \qquad i=1,2,3
\ee while the mean Hubble parameter is
\be\label{Hg}
H^{(g)}=\frac{1}{3}\sum_{i=1}^3{H_i^{(g)}}=H+\frac{\dot{B}}{B}-\frac{\dot{A}}{A}
\ee while for $\Delta H_i^{(g)}=H^{(g)}-H_i^{(g)}$, we obtain
\be
\Delta H_i^{(g)}=\Delta H_i, \qquad i=1,2,3
\ee

From the energy conservation of the matter, it follows that the thermodynamic parameters of the matter in the Universe must satisfy the $g$-metric conservation equation
\begin{equation}
\dot{\rho}+3H^{(g)}\left( \rho +p\right) =0 \label{cons}
\end{equation}
where $H^{(g)}$ is the Hubble parameter as defined in the anisotropic physical metric $g$. Taking into account Equation~(\ref{Hg}) in the $q$-space, the energy-momentum conservation equation takes the form
\be\label{consg}
\dot{\rho}+3\left(H+\frac{\dot{B}}{B}-\frac{\dot{A}}{A}\right)\left(\rho +p\right)=0
\ee

By assuming that the cosmological fluid obeys a barotropic equation of state so that $p=p(\rho )$, Equation~(\ref{consg}) can be integrated to provide
\be\label{VV}
\frac{\rho _0}{V}=\frac{B^3}{A^3}\exp\left[\int{\frac{d\rho }{\rho +p(\rho )}}\right]
\ee Therefore, the general solution of the field equations can be obtained in a parametric form, with $V$ taken as parameter, given by
\bea
&&a_i=a_{i0}V^{1/3}\times \exp \Bigg\{K_i \int{\frac{dV}{V\sqrt{C_0+(6/\kappa )\int{(1-1/AB)VdV}}}}\Bigg\}, \qquad i=1,2,3
\eea
\be H=\frac{1}{3V}\sqrt{C_0+\frac{6}{\kappa }\int{\left(1-\frac{1}{AB}\right)VdV}}
\ee and
\be A_p=\frac{1}{3}\frac{K^2}{V^2H^2}=\frac{3K^2}{\dot{V}^2}=\frac{3K^2}{\left(6/\kappa\right)\int{\left(1-1/AB\right)VdV}+C_0}
\ee The scale factors in the $g$-metric can be obtained as
\be \small{ g_i=\frac{B}{A}a_i=a_{i0}\frac{\sqrt{1-8\pi \kappa p}}{\sqrt{1+8\pi \kappa \rho} }V^{1/3} \exp \Bigg\{ K_i\int{\frac{dV}{V\sqrt{C_0+(6/\kappa )\int{(1-1/AB)VdV}}}}\Bigg\}, \qquad i=1,2,3}
\ee

For the comoving volume $V^{(g)}$ of the Universe we have
\be V^{(g)}=\frac{\left(1+8\pi \kappa p \right)^{3/2}}{\left(1-8\pi \kappa \rho \right)^{3/2}}V
\ee

Finally, for the anisotropy parameter $A_p^{(g)}$ in the physical metric $g$ we find
\be\label{an}
A_p^{(g)}=\frac{1}{3}\sum_{i=1}^{3}{\left[ \frac{\Delta H_{i}^{(g)}}{H^{(g)}}\right] ^{2}}=\frac{1}{3}\frac{K^2}{V^2\left(H+\dot{B}/B-\dot{A}/A\right)^2}
\ee

By substituting $H^{(g)}$ from the energy conservation Equation~(\ref{cons}) into Equation~(\ref{an}), we obtain
\be\label{anpar}
A_p^{(g)}=\frac{3K^2\left(\rho +p\right)^2}{V^2\dot{\rho}^2}
\ee

From Equation~(\ref{consg}) we obtain
\be
\frac{dV}{V}=\left\{-\frac{1}{\rho +p(\rho )}+12\pi \kappa \left[\frac{dp(\rho)}{d\rho }\frac{1}{B^2}+\frac{1}{A^2}\right]\right\}d\rho
\ee giving
\be VdV=\rho _0^2\frac{A^6}{B^6}\left\{-\frac{1}{\rho +p(\rho )}+12\pi \kappa \left[\frac{dp(\rho)}{d\rho }\frac{1}{B^2}+\frac{1}{A^2}\right]\right\}\exp\left[-2\int{\frac{d\rho }{\rho +p(\rho) }}\right]d\rho
\ee

\section{High Density Bianchi Type I Models in EiBI Gravity}\label{sect3}

In the following we will investigate the time dependence of the geometrical and thermodynamical parameters of the Bianchi type I space-times in the EiBI model for a number of equations of state that could be relevant for the description of the ultra-high density matter of which the Universe consisted in its very early stages.

\subsection{Stiff Fluid Filled Bianchi Type I Universe}

One of the most common equations of state, which has been used extensively to study the properties \scalebox{.93}{of the compact objects is the linear barotropic equation of state, with $p = (\gamma - 1)\rho $, with $\gamma = \mathrm{%
constant} \in [1,2]$}. The Zeldovich equation of state, valid for densities significantly higher than nuclear densities, $\rho > 10\rho _n$, can be obtained by constructing a relativistic Lagrangian that allows bare nucleons to interact attractively via scalar meson exchange, and repulsively via the exchange of a more massive vector meson \cite{60}. \mbox{In the non-relativistic} limit both the quantum and classical theories yield Yukawa-type potentials.
\mbox{At the highest} densities the vector meson exchange dominates and, by using a mean field approximation, \mbox{one can} show that in the extreme limit of infinite densities the pressure tends to the energy density, $p \rightarrow \rho $. In this limit the sound speed $c_s ^2=dp/d\rho \rightarrow 1$, and hence the stiff fluid equation of state satisfies the causality condition, with the speed of sound less than the speed of light.

For the specific case of $\rho =p$, the integral of the energy conservation Equation~(\ref{VV}) gives the comoving volume $V$ as a function of the density in the form \newpage
\begin{equation}\label{V1}
V(\rho )=\rho _0\frac{ (1+8 \pi \kappa \rho )^{3/2}}{\sqrt{\rho } (1-8 \pi \kappa
\rho )^{3/2}}
\end{equation}

For the comoving volume in the physical $g$ metric we obtain
\be V^{(g)}=\frac{\rho _0}{\sqrt{\rho }}
\ee which yields
\be
\rho =p=\frac{\rho _0^2}{\left(V^{(g)}\right)^2}
\ee

Taking into account Equation~(\ref{V1}), from Equation~(\ref{V}) we obtain the following equation for the time evolution of the density of the stiff fluid filled Bianchi type I Universe,
\begin{eqnarray}\label{dens1}
&&2\kappa \rho \left( 4096\pi ^{4}\kappa ^{4}\rho ^{4}+3072\pi ^{3}\kappa
^{3}\rho ^{3}-128\pi ^{2}\kappa ^{2}\rho ^{2}-48\pi \kappa \rho +1\right)
\ddot{\rho }
\nonumber \\
&&-3\kappa \left( 4096\pi ^{4}\kappa ^{4}\rho
^{4}+6144\pi ^{3}\kappa ^{3}\rho ^{3}+
640\pi ^{2}\kappa ^{2}\rho ^{2}-32\pi
\kappa \rho +1\right) \dot{\rho }^{ 2}
\nonumber \\
&&
-12\rho ^{2}\left( 1-64\pi ^{2}\kappa ^{2}\rho ^{2}\right) ^{3/2}\left[ 1-\left( 1-64\pi ^{2}\kappa ^{2}\rho ^{2}\right) ^{1/2}\right] =0
\end{eqnarray}

By performing a series expansion with respect to the density, from Equation~(\ref{V1}) we obtain for the comoving volume in the $q$ metric the following expression
\begin{equation}\label{exp1}
V\left(\rho \right)\approx \frac{\rho _0}{\sqrt{\rho }}+24 \pi \kappa \rho _0 \sqrt{\rho
}+288 \pi ^2 \kappa ^2 \rho _0 \rho ^{3/2}+2816 \pi ^3 \kappa ^3
\rho _0 \rho ^{5/2}+O\left(\rho ^{7/2}\right)
\end{equation} \vspace{-6pt}

\subsubsection{The First Order Approximation}

In the first order approximation we keep only the first term in the right hand side of Equation~(\ref{exp1}). Hence we obtain for the comoving volume the simple relation $V\approx \rho _0/\sqrt{\rho }$. In this approximation, Equation~(\ref{V}) takes the form
\be\label{32n}
\ddot{V}=\frac{3}{\kappa }\left(1-\frac{1}{\sqrt{1-64\pi ^2\kappa ^2\rho _0^4/V^4}}\right)V
\ee

A first integration of Equation~(\ref{32n}) gives
\be
\dot{V}^2=C_0+\frac{3}{\kappa }V^2 \left(1-\sqrt{1-\frac{64 \pi ^2 \kappa ^2 \rho _0^2}{V^4}}\right)
\ee

In order to have a physical model defined for all times, the comoving volume must satisfy the condition $V=V_0\geq 8\pi \kappa \rho _0$. Therefore in EiBI gravity, the stiff causal Universe starts its evolution from a non-singular state, with the initial density $\rho _{in}$ given by $\rho _{in}=\rho _0/V_0^2$.

Hence in the first order approximation, the time dependence of the evolution of the comoving volume in the $q$-metric is given by\newpage
\be t-t_0=\int{\frac{dV}{\sqrt{C_0+\frac{3}{\kappa }V^2 \left(1-\sqrt{1-\frac{64 \pi ^2 \kappa ^2 \rho _0^4}{V^4}}\right)}}}
\ee while the time evolution of the physical scale factors is obtained as
\bea a_i=a_{i0}\frac{\sqrt{1-8\pi \kappa \rho _0/V^2}}{\sqrt{1+8\pi \kappa \rho _0}}V^{1/3}
\exp\left[K_i\int{\frac{dV}{V\sqrt{C_0+\frac{3}{\kappa }V^2 \left(1-\sqrt{1-\frac{64 \pi ^2 \kappa ^2 \rho _0^4}{V^4}}\right)}}}\right]
\eea

For the scale factors in the physical metric $g$ we obtain {\small
\bea g_i=a_{i0}\frac{\sqrt[3]{\rho _0}}{2 \sqrt[6]{\rho }} \exp \left\{K_i\int \frac{1-16 \pi \kappa \rho (4 \pi \kappa
\rho +3)}{\rho \left(64 \pi ^2 \kappa ^2 \rho ^4-1\right) \sqrt{C_0+\frac{3
\rho _0^4 \left[16 \pi \kappa \rho (4 \pi \kappa \rho -3)-35
\left(\sqrt{1-64 \pi ^2 \kappa ^2 \rho ^2}-1\right)\right] (8 \pi \kappa \rho
+1)^3}{35 \kappa \rho (8 \pi \kappa \rho -1)^3 \sqrt{1-64 \pi ^2 \kappa ^2
\rho ^4}}}} \, d\rho \right\}
\eea
}
for $i=1,2,3$. The mean value of the Hubble parameter is found as
\bea H^{(g)}=\frac{1}{3V}\sqrt{C_0+\frac{3}{\kappa }V^2 \left(1-\sqrt{1-\frac{64 \pi ^2 \kappa ^2 \rho _0^4}{V^4}}\right)}
\left[1+\frac{48\pi \kappa \rho _0}{V^2\left(1-8\pi \kappa \rho _0/V^2\right)\left(1+8\pi \kappa \rho _0/V^2\right)}\right]
\eea

The time evolution of the anisotropy parameter is described by the equation
\begin{equation}
A_{p}=\frac{3K^{2}}{C_{0}+\left( 3/\kappa \right) V^{2}\sqrt{1-64\pi
^{2}\kappa ^{2}\rho _{0}^{4}/V^{4}}}
\end{equation}\vspace{-6pt}

\subsubsection{General Dynamics of the Stiff Fluid Filled Bianchi Type I Universe}

\begin{figure}[H]
\centering
\includegraphics[width=8.65cm]{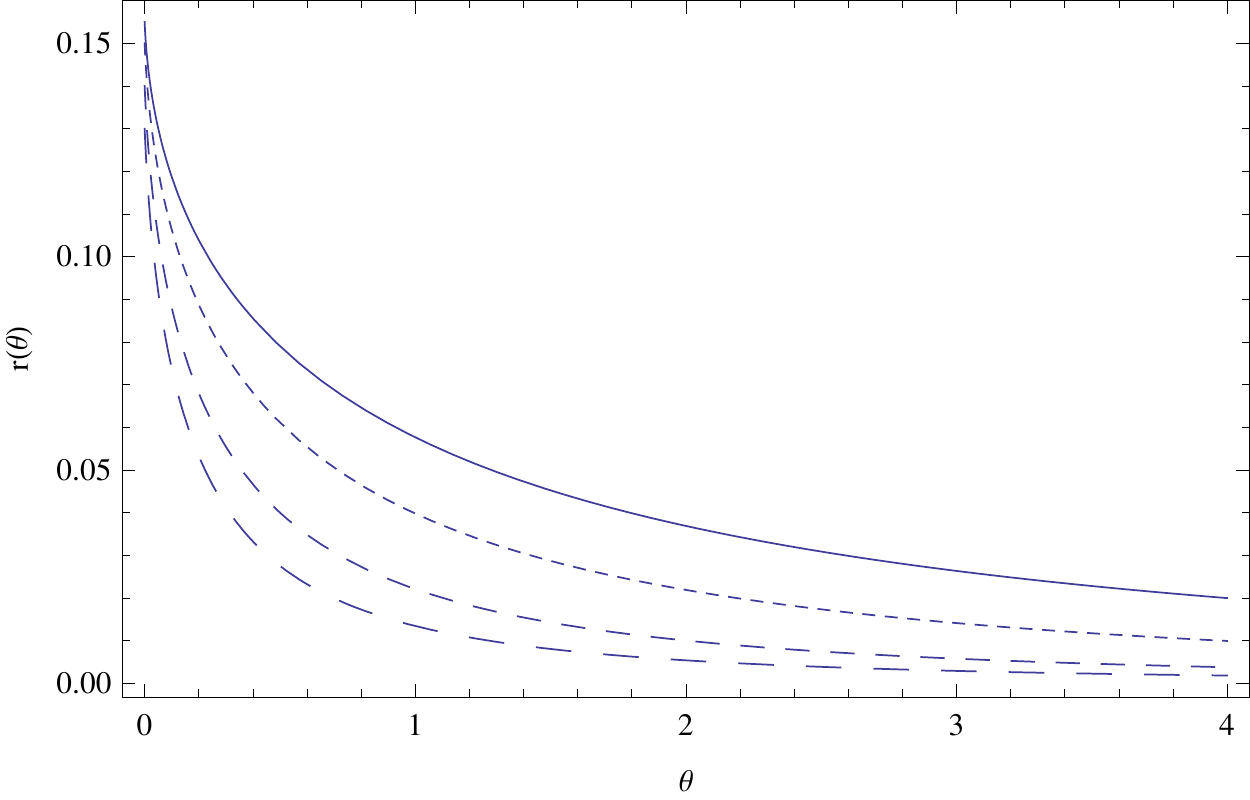}
\includegraphics[width=8.5cm]{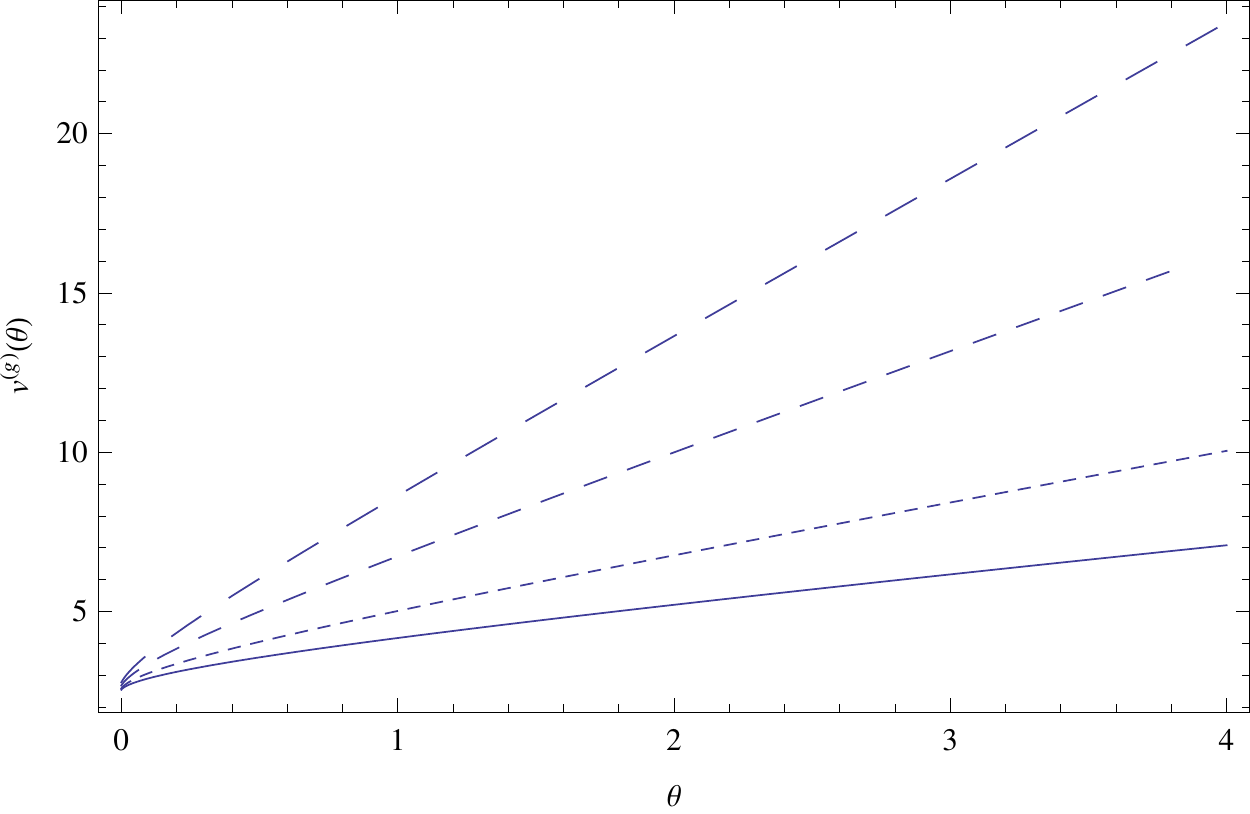}
\caption{Variation of the dimensionless energy density $r$ of the stiff fluid filled Bianchi type I space-time in EiBI gravity (\textbf{left} figure) and of the comoving volume $v^{(g)}$ in the physical space $g$ (\textbf{right} figure), for different initial values of the dimensionless energy density \mbox{$r$: $r(0)=0.155$} (solid curve), $r(0)=0.15$ (dotted curve), $r(0)=0.14$ (short dashed curve) and $r(0)=0.13$ (dashed curve), respectively. In all cases $r'(0)=-1.5$.}
\label{fig1}
\end{figure}

In order to study the exact evolution of the stiff fluid filled Bianchi I type universe in EiBI gravity, \mbox{we introduce} first a set of dimensionless parameters $\left(\theta,r\right)$, defined as
\begin{equation}\label{dvar}
\theta=\frac{t}{\sqrt{\kappa }}, \qquad \rho =\frac{r}{8\pi \kappa }
\end{equation}
In these variables Equation~(\ref{dens1}) takes the form
\bea\label{r1}
-2 \left(r^4+6 r^3-2 r^2-6 r+1\right) r r''+3 \left(r^4
+12 r^3+10 r^2-4 r+1\right) r'^2
\nonumber\\
-12 \left(r^2-1\right)^2 r^2- 12
\sqrt{1-r^2} \left(r^2-1\right) r^2=0
\eea where a prime denotes the derivative with respect to $\theta$. The energy density must satisfy the constraint $r\leq 1$. The time variation of the energy density of the stiff fluid, and of the comoving volume in \mbox{the physical} $g$ space, $v^{(g)}=V^{(g)}/\rho _0\sqrt{8\pi \kappa }$, is represented in Figure~\ref{fig1}.

By introducing the rescaled anisotropy parameter $a_p^{(g)}$ in the physical $g$-space, defined as
\be a_p^{(g)}=\frac{\rho _0^2}{96\pi \kappa K^2}A_p^{(g)}
\ee we obtain
\be a_p^{(g)}=\frac{r^3(1+r)^3}{(1-r)^3r'^2}
\ee

\begin{figure}[H]
\centering
\includegraphics[width=8.4cm]{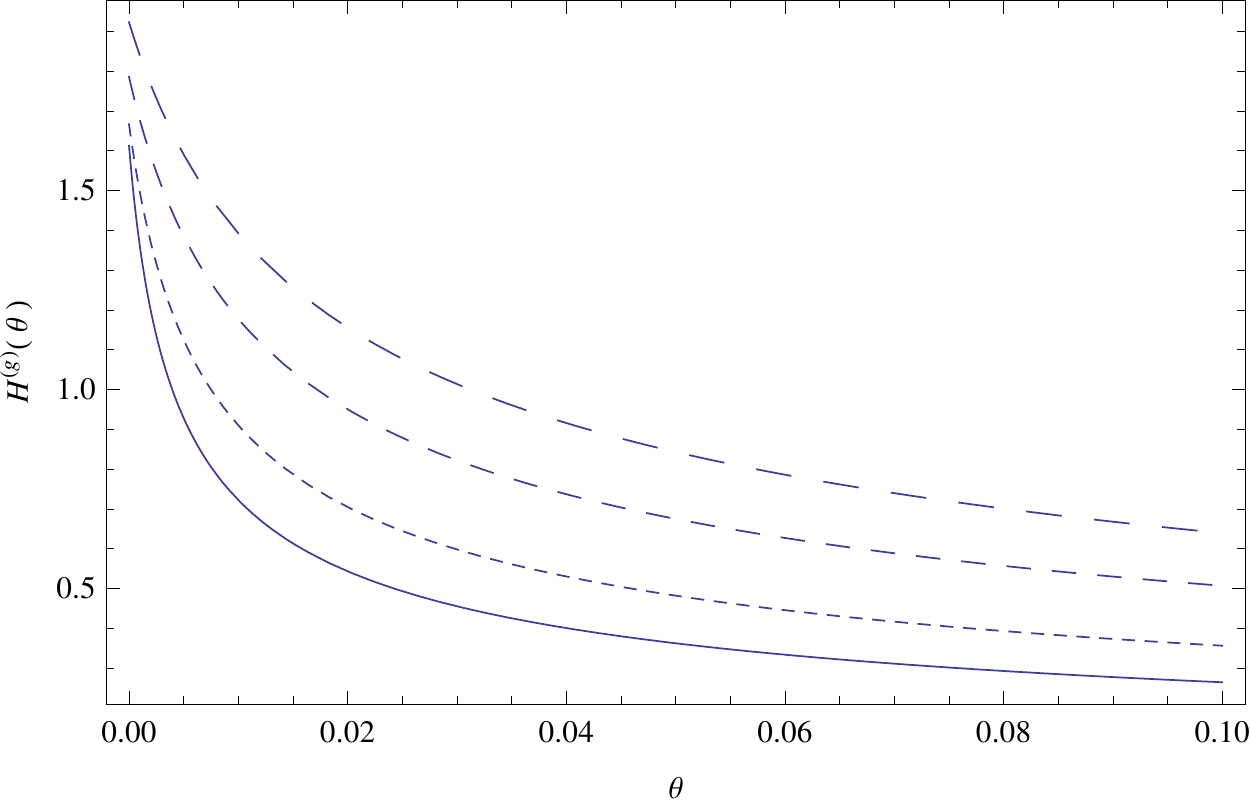}
\includegraphics[width=8.55cm]{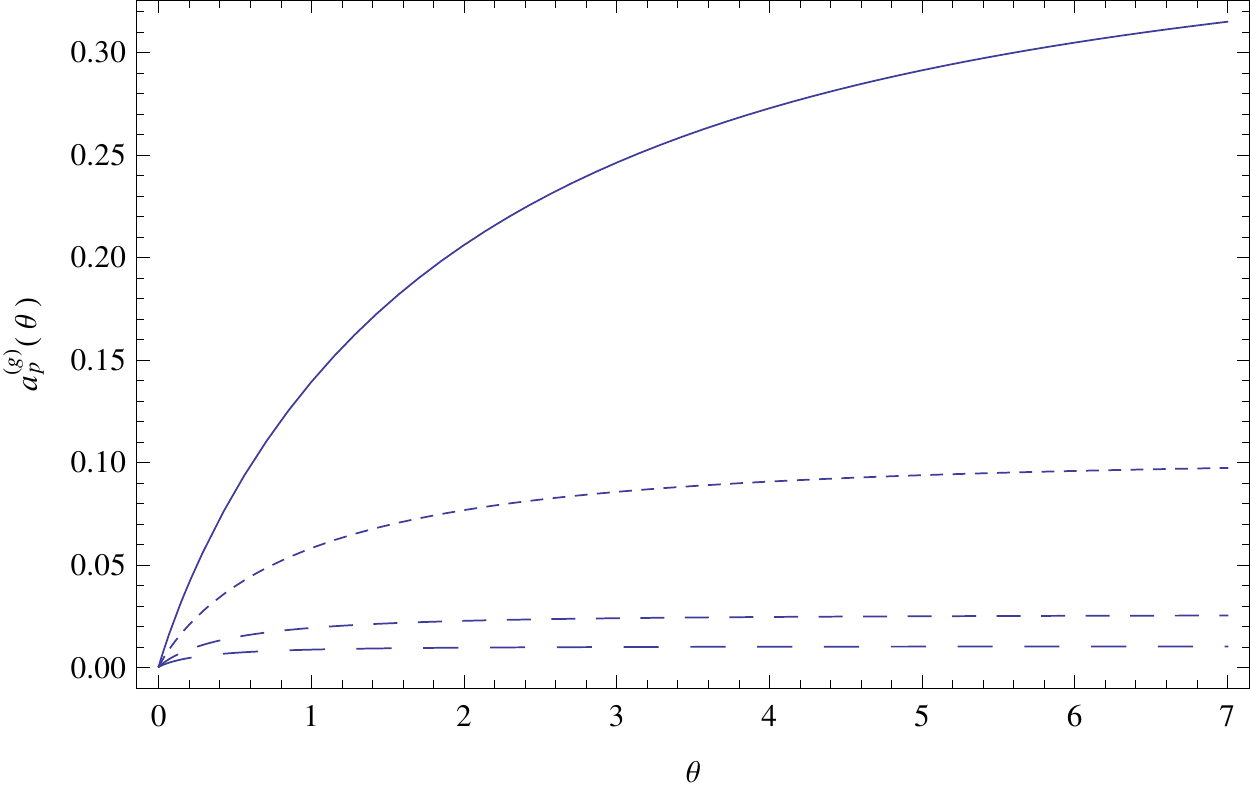}
\caption{Variation of the dimensionless Hubble function $H^{(g)}$ of the stiff fluid filled Bianchi type I space-time in EiBI gravity (\textbf{left} figure) and of the mean anisotropy parameter $a^{(g)}_p$ in the physical space $g$ (\textbf{right} figure), for different initial values of the dimensionless energy density $r$: $r(0)=0.155$ (solid curve), $r(0)=0.15$ (dotted curve), $r(0)=0.14$ (short dashed curve) and $r(0)=0.13$ (dashed curve), respectively. As before, we have considered $r'(0)=-1.5$ for all cases. }
\label{fig2}
\end{figure}

The time variation of the Hubble function and of the mean anisotropy parameter are represented \mbox{in Figure~\ref{fig2}.}

For the chosen initial conditions during the entire anisotropic cosmological evolution, the matter density of the stiff fluid filled Universe is a monotonically decreasing function of time, while the volume element of the anisotropic space-time increases linearly. The Hubble function is a monotonically decreasing function of time, and in the large time limit it tends to zero, $\lim_{\theta \rightarrow \infty}H^{(g)}(\theta )=0$. On the other hand, the anisotropy parameter is an increasing function of time, starting from a zero initial value. \mbox{The Universe} is born in isotropic state, with the anisotropy parameter being zero. Then, during the cosmological evolution, its degree of anisotropy increases gradually. In the large time limit, the Universe ends in a state of constant anisotropy, so that $\lim _{\theta \rightarrow \infty }a_p^{(g)}={\rm constant}$. The numerical value of the anisotropy parameter in the large time limit is strongly dependent on the initial conditions of\linebreak the density.

\subsection{The Radiation Fluid}

As a second example of a Bianchi type I Universe in EiBI gravity, we consider the case in which the matter content of the Universe consists of a radiation type fluid, with equation of state given by
\be p=\frac{\rho }{3}
\ee

Then from the matter conservation Equation~(\ref{VV}) we obtain the volume element in the $q$-space as \mbox{a function} of the matter energy density in the form
\begin{equation}
V=\rho _{0}\frac{\left( 1+8\pi \kappa \rho \right) ^{3/2}}{\rho ^{3/4}\left( 1-8\pi \kappa \rho /3\right) ^{3/2}}
\end{equation}

\scalebox{.98}{With the use of the dimensionless variables introduced in Equation~(\ref{dvar}), the comoving volume becomes}
\begin{equation}
V=\rho _{0}\left( 8\pi \kappa \right) ^{3/4}\frac{\left( 1+r\right) ^{3/2}}{%
r^{3/4}\left( 1-r/3\right) ^{3/4}}
\end{equation}

Equation (\ref{VV}) gives the time evolution of the dimensionless density $r$ as the following second order nonlinear differential equation,
\bea 36\sqrt{3}\left( r^{4}+4r^{3}-18r^{2}-12r+9\right) rr^{\prime \prime }-9%
\sqrt{3}\left( 7r^{4}+84r^{3}+18r^{2}-60r+63\right) r^{\prime 2}
\nonumber \\
+144\sqrt{3}%
\left( r^{2}-2r-3\right) \left( r^{2}-2r+\sqrt{-3r^{2}+6r+9}-3\right) r^{2}=0
\eea

In order to obtain real values for the energy density, $r$ must satisfy the constraint $r\leq 3$. The comoving volume in the physical space $g$ is obtained as
\be V^{(g)}=\frac{\rho _0}{\rho ^{3/4}}=\frac{\rho _0\left(8\pi \kappa\right)^{3/4}}{r}
\ee or, by introducing the rescaled volume element $v^{(g)}=V^{(g)}/\rho _0\left(8\pi \kappa\right)^{3/4}$, as
\be v^{(g)}=\frac{1}{r^{3/4}}
\ee

\begin{figure}[H]
\centering
\includegraphics[width=8.5cm]{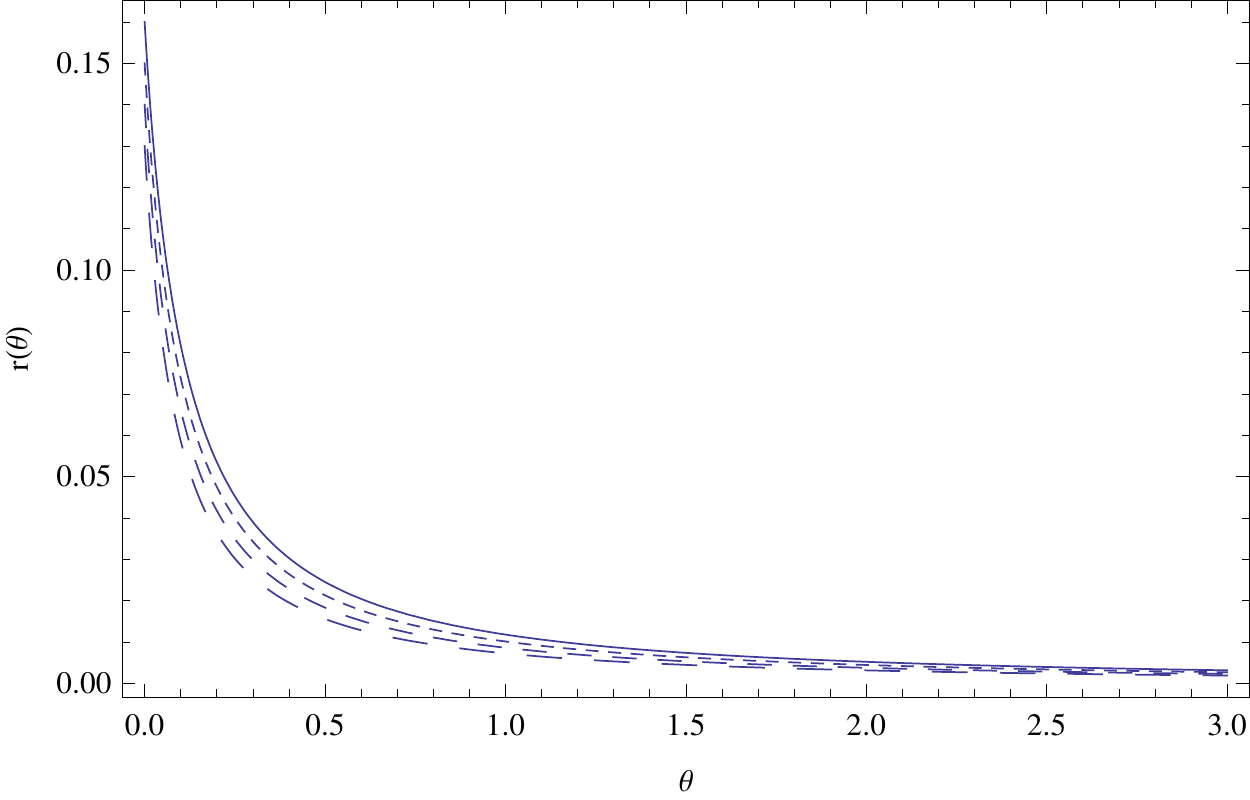}
\includegraphics[width=8.5cm]{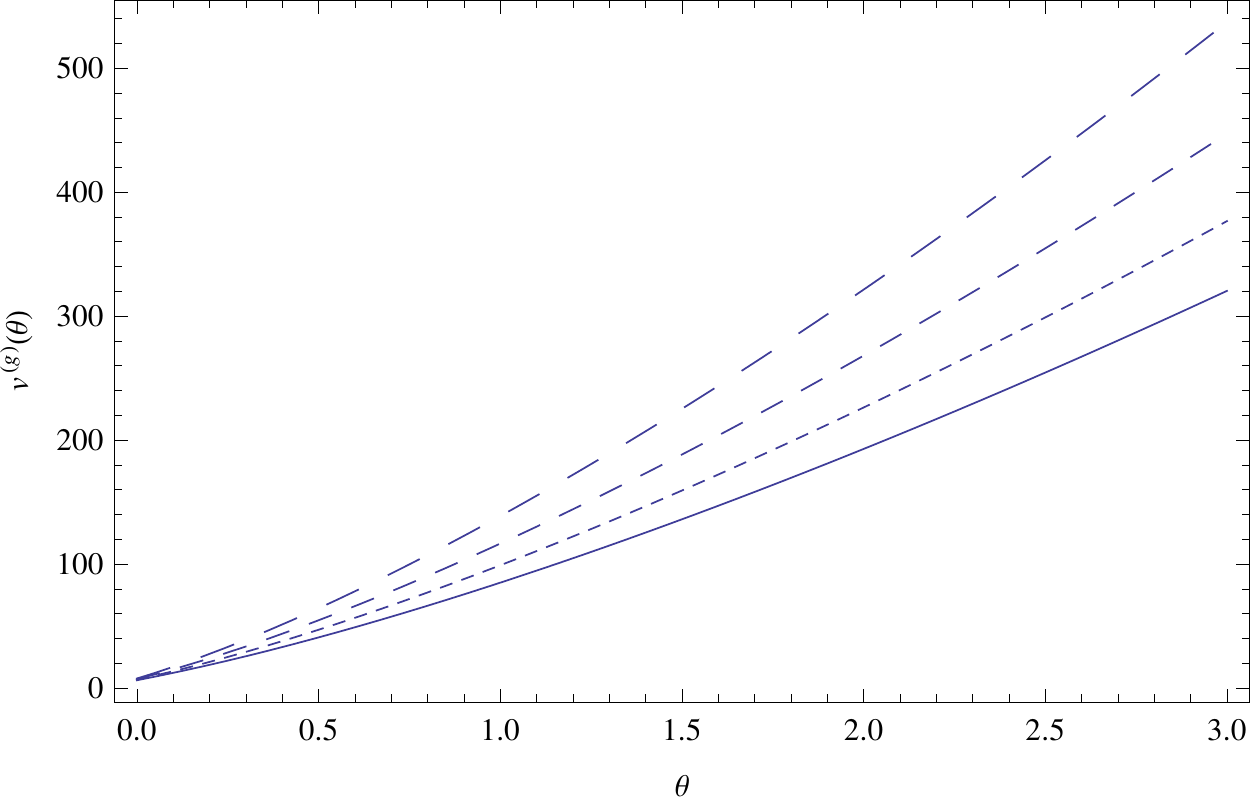}
\caption{Time variation of the dimensionless energy density $r$ of the radiation filled Bianchi type I space-time in EiBI gravity (\textbf{left} figure) and of the comoving volume $v^{(g)}$ in the physical space $g$ (\textbf{right} figure), for different initial conditions of the density: $r(0)=0.16$ \mbox{(solid curve),} $r(0)=0.15$ (dotted curve), $r(0)=0.14$ (short dashed curve) and $r(0)=0.13$ (dashed curve), respectively. In all cases $r'(0)=-1.5$. }
\label{fig1r}
\end{figure}

The variations of the energy density and of the rescaled volume element in the physical space are represented in Figure~\ref{fig1r}.

\begin{figure}[H]
\centering
\includegraphics[width=8.1cm]{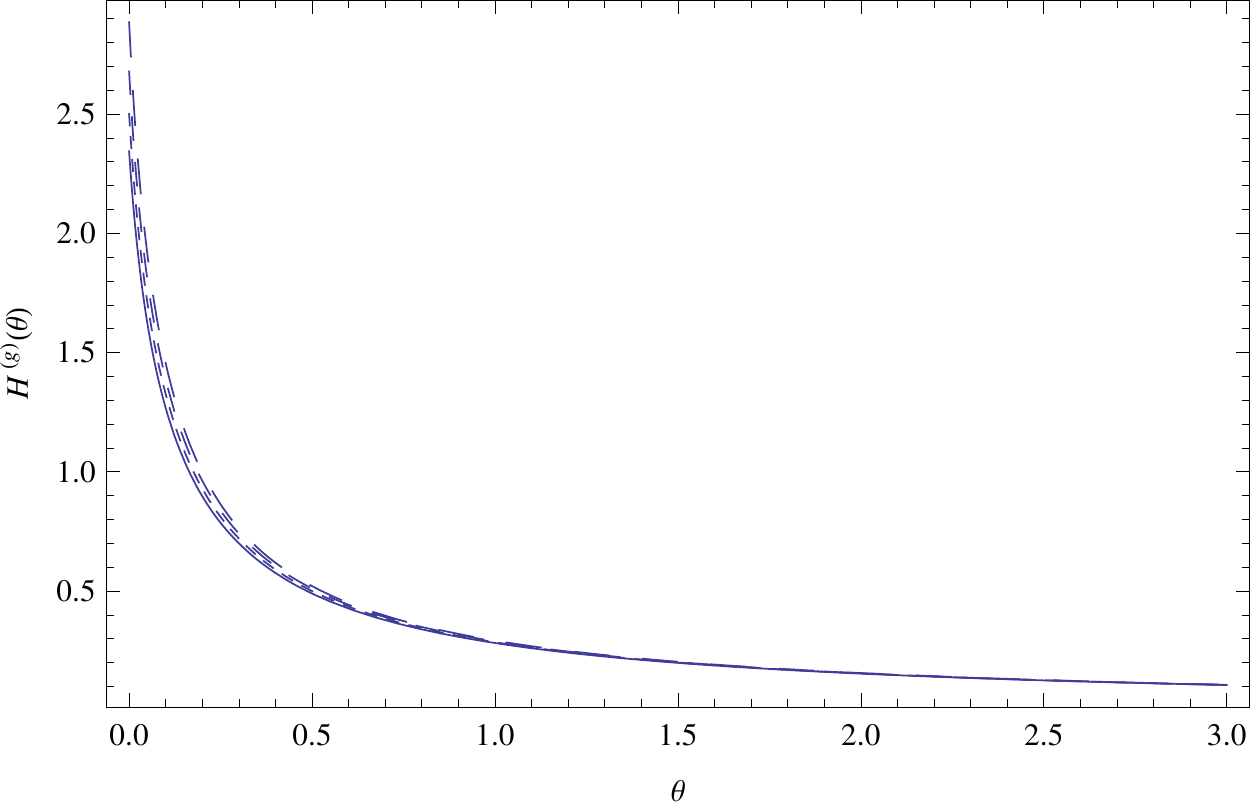}
\includegraphics[width=8.5cm]{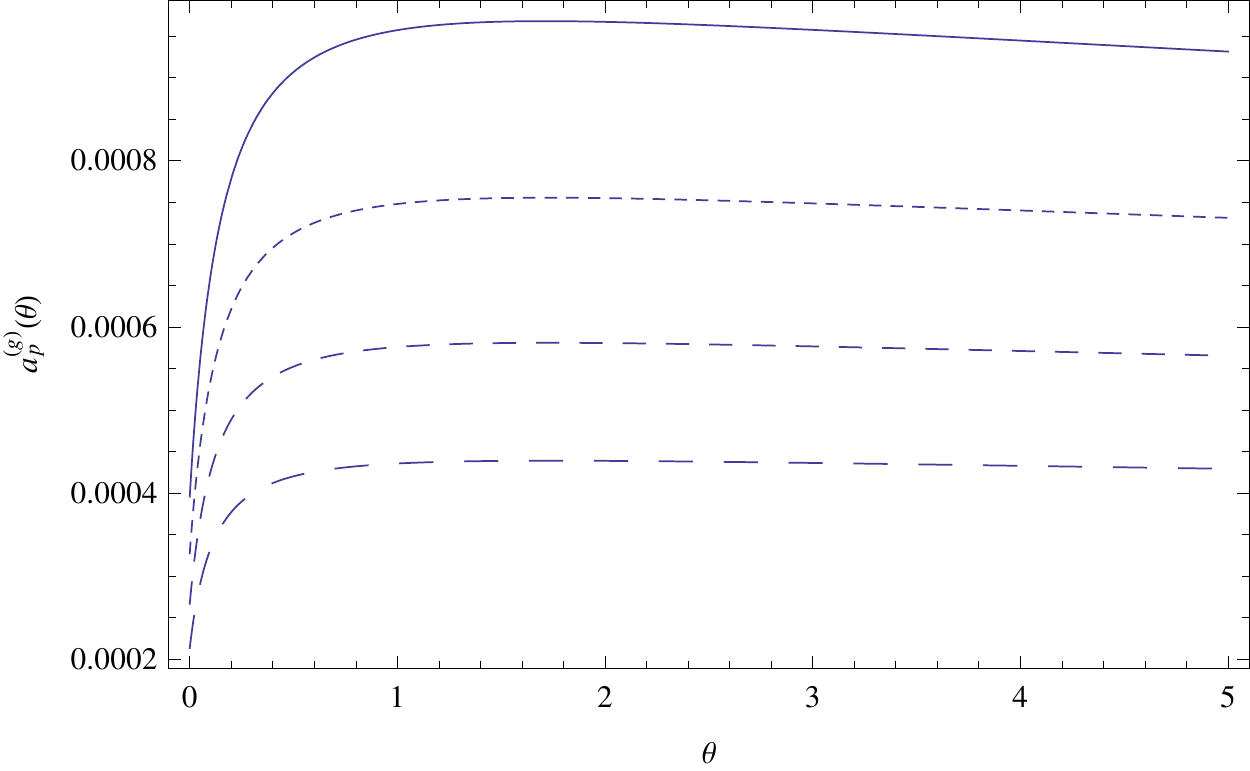}
\caption{Time variation of the physical Hubble function $H^{(g)}$ of the radiation filled Bianchi type I space-time in EiBI gravity (\textbf{left} figure) and of the rescaled mean anisotropy parameter $a^{(g)}_p$ in the physical space $g$ (\textbf{right} figure), for different initial conditions of the density: \mbox{$r(0)=0.16$} (solid curve), $r(0)=0.15$ (dotted curve), $r(0)=0.14$ (short dashed curve) and $r(0)=0.13$ (dashed curve), respectively. In all cases $r'(0)=-1.5$. }
\label{fig2r}
\end{figure}

The time variations of the Hubble function in the physical space $H^{(g)}$, and of the rescaled anisotropy parameter $a_p^{(g)}$,
\be a_p^{(g)}=\rho _0^2\frac{\left(8\pi \kappa \right)^{3/4}}{48K^2}A_p^{(g)}=\frac{(1-r/3)^3r^{7/2}}{(1+r)^3r'^2}
\ee are presented in Figure~\ref{fig2r}.

For the adopted initial conditions the behavior of the radiation filled Bianchi type I Universe in EiBI gravity is qualitatively similar to the dynamics in the stiff fluid case. The energy density and the Hubble function are monotonically decreasing functions of time, while the comoving volume in the physical space is increasing during the cosmological evolution. The anisotropy parameter increases from a small initial value to a maximum constant value, which strongly depends on the initial conditions of \mbox{the energy density.}

\section{Pressureless Anisotropic Bianchi Type I Universes in EiBI Gravity}\label{sect4}

In the case of dust matter, $p=0$, and we have $B=1$. Then the energy conservation Equation~(\ref{VV}) gives the following relation for the time variation of the energy density in the dust Bianchi \linebreak type I Universe:
\be V=\frac{\rho _0}{\rho \left(1+8\pi \kappa \rho \right)^{3/2}}
\ee

\subsection{The First Order Approximation}

In the first order approximation, by assuming that the energy density of the Universe is low, so that $8\pi \kappa \rho \ll 1$, for $\rho $ we obtain,
\begin{equation}
\rho =\frac{\rho _0}{V}
\end{equation}

By series expansion we can express the functions $A$ and $1/A$ as
\be A=\sqrt{1+8\pi \kappa \rho }\approx 1+\frac{4\pi \kappa \rho _0}{V},
\qquad
\frac{1}{A}\approx 1-\frac{4\pi \kappa \rho _0}{V}
\ee

By first order approximation, Equation~(\ref{V}) becomes
\begin{equation}
\ddot{V}=12\pi \rho _0
\end{equation}
with the general solution given by
\begin{equation}
V(t)=C_2+C_1t+6\pi \rho _0 t^2
\end{equation}
where $C_1$ and $C_2$ are arbitrary constants of integration. In order to obtain the explicit form of the integration constants, we use the initial conditions $V(0)=V_0$ and $\dot{V}(0)=3V_0H_0$, where $H_0=H(0)$. Hence we obtain
\be C_2=V_0, \qquad C_1=3V_0H_0
\ee

The Hubble parameter is found as
\begin{equation}
H=\frac{V_0H_0+4\pi \rho _0t}{V_0+3V_0H_0t+6\pi \rho _0 t^2}
\end{equation}
while the time evolution of the scale factors is given, as a function of $V$, as
\be a_i=a_{i0}V^{1/3}\exp\left[-\frac{2 K_i \tanh ^{-1}\left(\frac{\sqrt{C_0+24 \kappa \rho
_0 V}}{\sqrt{C_0}}\right)}{\sqrt{C_0}}\right], \qquad i=1,2,3
\ee or, equivalently,
\begin{eqnarray}
a_i=a_{i0} \sqrt[3]{V_0+3V_0H_0t+6\pi \rho _0 t^2}
\exp \left[-\frac{2 K_i \tanh ^{-1}\left(\frac{\sqrt{C_0+24 \kappa
\rho _0 \left(3 H_0 t V_0+6 \pi \rho _0 t^2+V_0\right)}}{\sqrt{C_0}}\right)}{\sqrt{C_0}}\right]
\end{eqnarray}
with $i=1,2,3$. For the mean value of the anisotropy parameter we obtain
\be A=\frac{3K^2}{\left(3V_0H_0+12\pi \rho _0t\right)^2}
\ee

In the limit of large times the anisotropy parameter tends to zero, $\lim_{t\rightarrow \infty}A=0$, showing that the dust Bianchi type I Universe will end its evolution in isotropic phase.

\subsection{Evolution of the Dust Bianchi Type I Universe}

With the use of the dimensionless variables introduced through Equation~(\ref{dvar}), the comoving volume $V$ in the $q$-space can be expressed as
\be V=8\pi \kappa \rho _0\frac{1}{r\left(1+r\right)^{3/2}}
\ee

Equation (\ref{V}) gives then the time variation of the density of the dust Bianchi type I Universe in EiBI gravity as
\be 2(r+1)(5r+2)rr^{\prime \prime }-\left[ 7r(5r+4)+8\right] r^{\prime 2}-12%
\sqrt{r+1}r^{2}\left[ r-\left( 1+r\right) ^{3/2}+1\right] =0
\ee

In addition to this, for the comoving volume $V^{(g)}$ in the physical space $g$ we obtain
\be V^{(g)}=\frac{\rho _0}{\rho }=\frac{8\pi \kappa \rho _0}{r}
\ee

\begin{figure}[H]
\centering
\includegraphics[width=7.8cm]{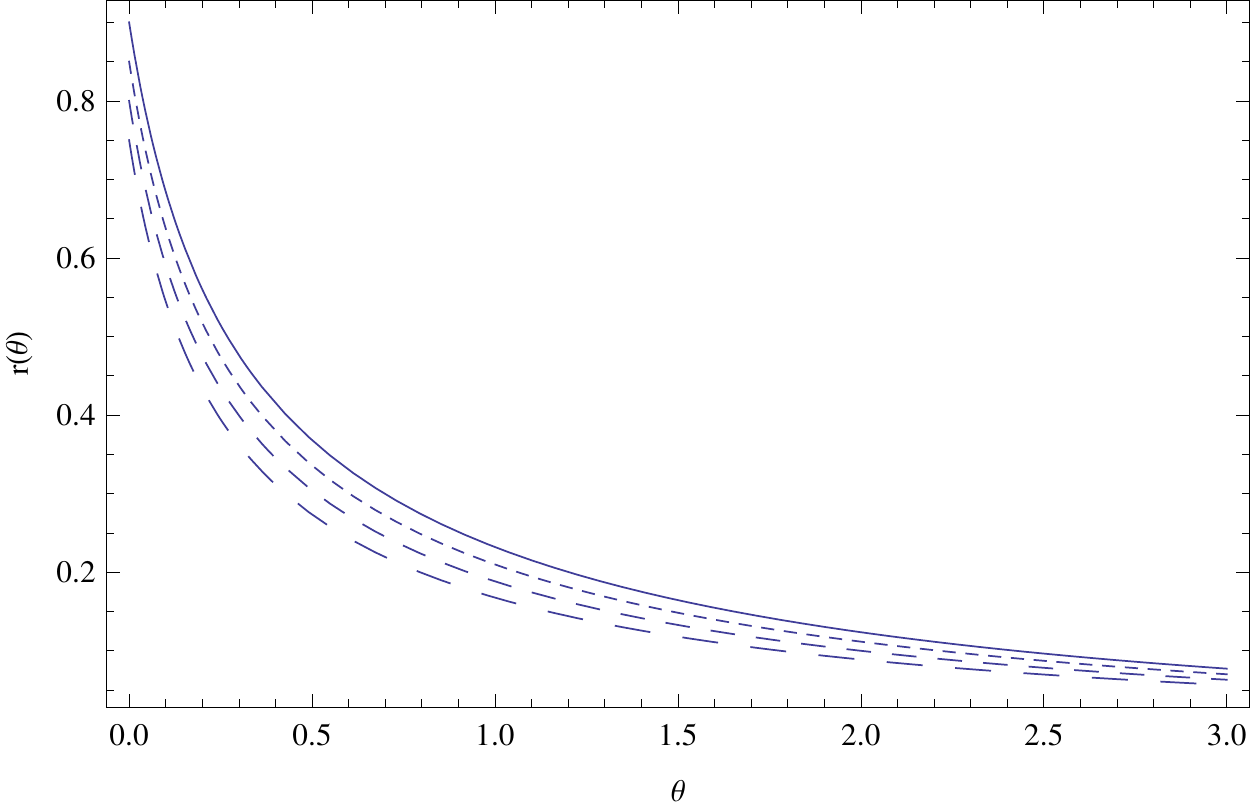}
\includegraphics[width=7.8cm]{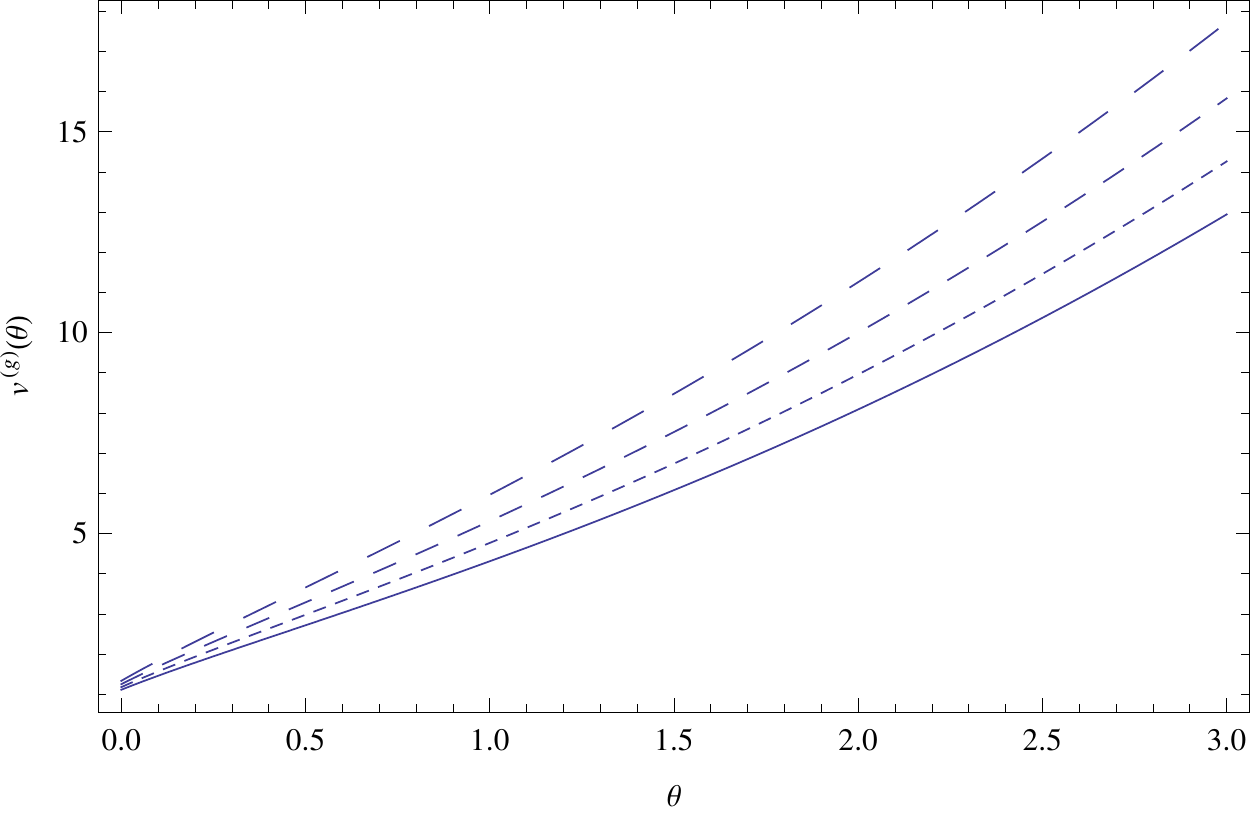}
\caption{Time variation of the dimensionless matter density $r$ of the dust fluid filled Bianchi type I space-time in EiBI gravity (\textbf{left} figure) and of the scaled volume element $v^{(g)}$ in the physical space $g$ (\textbf{right} figure), for different initial conditions of the density: $r(0)=0.9$ (solid curve), $r(0)=0.85$ (dotted curve), $r(0)=0.80$ (short dashed curve) and $r(0)=0.75$ (dashed curve), respectively. In all cases $r'(0)=-3$. }
\label{fig3}
\end{figure}

In this context, the time variations of the dimensionless energy density $r$ and of the scaled \mbox{volume element}
\be v^{(g)}=\frac{V^{(g)}}{8\pi \kappa \rho _0}
\ee are represented in Figure~\ref{fig3} for different initial conditions of the density.

\begin{figure}[H]
\centering
\includegraphics[width=7.8cm]{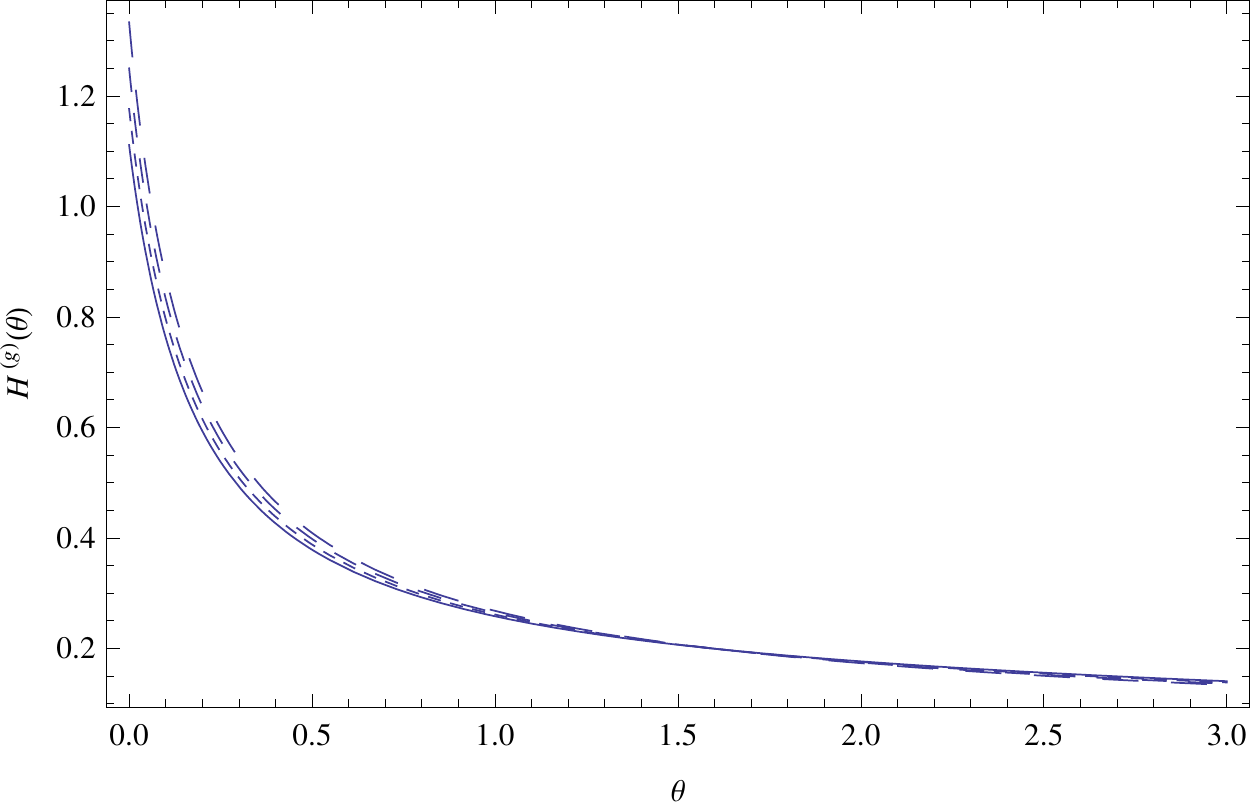}
\includegraphics[width=7.8cm]{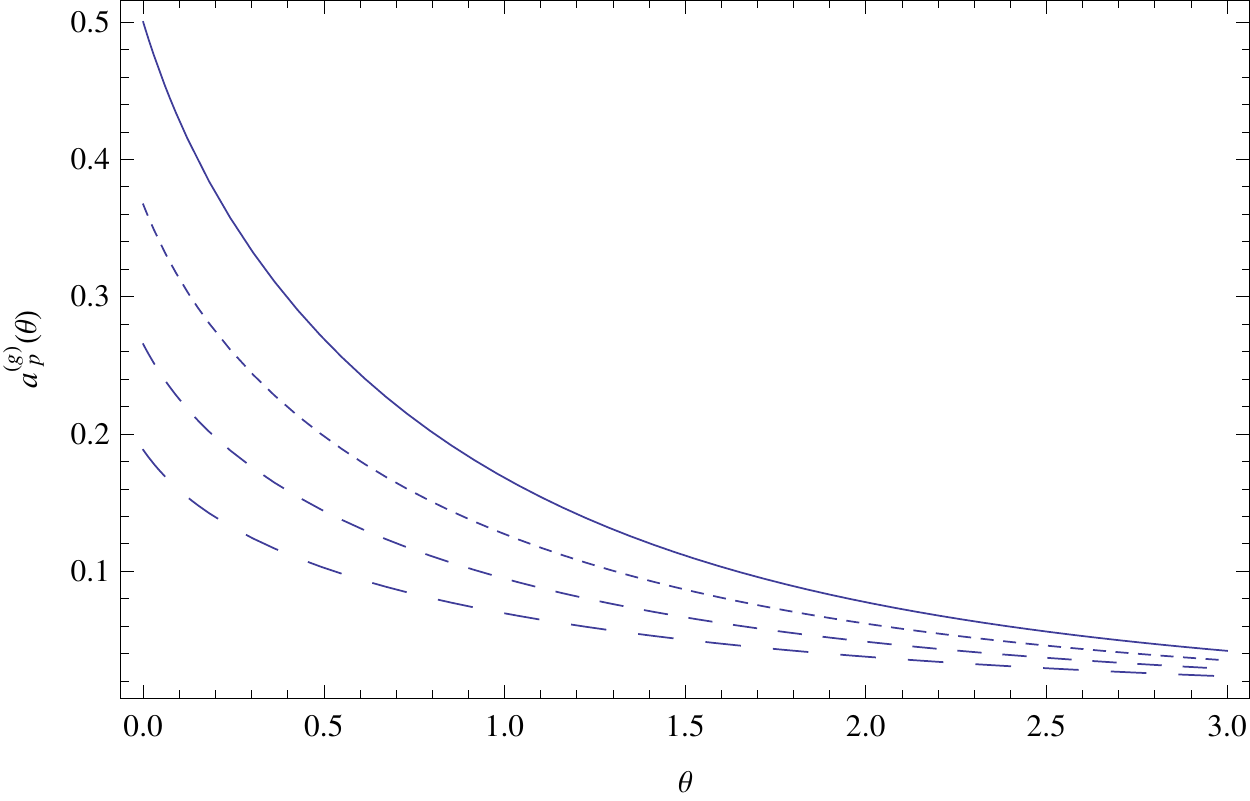}
\caption{Time variation of the dimensionless Hubble function $H^{(g)}$ of the dust fluid filled Bianchi type I space-time in EiBI gravity (\textbf{left} figure), and of the mean anisotropy parameter $a^{(g)}_p$ in the physical space $g$ (\textbf{right} figure), for different initial conditions of the density: $r(0)=0.9$ (solid curve), $r(0)=0.85$ (dotted curve), $r(0)=0.80$ (short dashed curve) and $r(0)=0.75$ (dashed curve), respectively. In all cases $r'(0)=-3$. }
\label{fig4}
\end{figure}

The time variations of the Hubble function and of the rescaled mean anisotropy parameter $a_p^{(g)}$ in the physical space $g$, defined as
\be a_p^{(g)}=\frac{\left(8\pi \kappa \rho _0\right)^2}{3K^2}A_p^{(g)}=\frac{r^4\left(1+r\right)^3}{r'^2}
\ee are represented in Figure~\ref{fig4}.

\vspace{-12pt}

In the long time limit the mean anisotropy parameter tends to zero, showing that in the EiBI gravity, a Bianchi type I dust filled Universe ends its evolution in isotropic phase. The energy density is a monotonically decreasing function of time, while the comoving volume element is increasing during the entire time interval of the cosmological expansion.

\section{Discussions and Final Remarks}\label{sect5}

In the present paper we have considered the cosmological evolution of Bianchi type I Universes in the Eddington-inspired Born--Infeld gravity theory, a two metric theory, based on the Palatini formalism, in which the metric $g_{\mu \nu}$ and the connection $\Gamma ^{\alpha }_{\beta \gamma }$ are assumed to be independent fields. The Ricci tensor $R_{\mu \nu}(\Gamma)$ is evaluated by the connection only, while the matter fields are coupled to the gravitational field via the metric $g_{\mu \nu}$ only. By adopting the homogeneous and anisotropic Bianchi type I geometry, a natural extension to the anisotropic case of the flat Friedmann--Robertson--Walker geometry, we have obtained, as a first step in our study, the gravitational field equations in the auxiliary $q$ metric. Under the supplementary assumption of isotropic pressure distribution, a standard choice in the study of anisotropic cosmological models, the basic equations of the model in the $q$ metric have been derived, and the representation of the physical and geometrical quantities in the $g$ metric has also been presented.

As concrete cosmological applications, we have investigated three distinct models, corresponding to different equations of state of the cosmological matter. As a first case, we have considered that the equation of state of the matter satisfies the stiff fluid equation of state, in which the energy density equals the thermodynamic pressure. The basic evolution equation in EiBI gravity is given by a complicated nonlinear second order differential equation for the density, Equation~(\ref{dens1}). This equation, as well as the general properties of the model, impose the constraint $\rho \leq c^2/8\pi \kappa G$, or, equivalently,
\be
\rho \leq 5.36\times 10^{26}\times \left(\frac{\kappa}{{\rm cm ^2}}\right)^{-1}\;{\rm g/cm^3}
\ee

By assuming for $\kappa $ a value in the order of $\kappa \approx 10^{12}\;{\rm cm ^2}$ \cite{prop1f}, as suggested by the study of massive compact objects in the EiBI theory, we obtain for the density of the stiff matter fluid the constraint $\rho \leq 5\times 10^{14}\;{\rm g/cm^3}$.

In order to study Equation~(\ref{dens1}) describing the time evolution of the matter density, one must give the initial conditions for the density, that is, its initial value at $t=0$, $\rho (0)$, and the initial value of its derivative, $\dot{\rho }(0)=-6H^{(g)}(0)\rho (0)$. The initial value of the density derivative is determined by the initial values of the Hubble function and of the density. Depending on the concrete numerical values of the density and of its derivative, a large number of different cosmological scenarios can be obtained. In the present paper, we have presented only one such model, which has the intriguing property that despite considering the expansion of the Universe, with the energy density and the Hubble function monotonically decreasing in time, the mean anisotropy parameter in the physical $g$ space is increasing from an initial small value and in the long time limit tends to a constant value. The behavior of the radiation fluid filled Bianchi type I Universe is very similar to the stiff fluid case, with the cosmological dynamics described by a strongly nonlinear second order differential equation, and an evolution that is dependent on initial conditions. The allowed range of the matter densities are restricted by the conditions $\rho \leq 1.61\times 10^{27}/ \left(\kappa /{\rm cm^2}\right)\;{\rm g/cm^3}$, or $\rho \leq 1.61\times 10^{15}\;{\rm g/cm^3}$. For initial conditions for the density similar to the stiff fluid case, we obtain a very similar dynamics, with the anisotropy parameter increasing in time in the initial phases of the cosmological evolution and reaching a constant value in the large time limit. These results suggest that the possible large scale anisotropies detected by Planck \cite{Plancka,Planckb,Planckc} may be due to the cosmological evolution in an initially anisotropic Bianchi type I Universe described by the Eddington-inspired Born--Infeld gravity theory. Depending on the assumed initial conditions, initially anisotropic Bianchi type I Universes may not end their cosmological evolution in isotropic and homogeneous flat Robertson--Walker type geometry.

{In standard general relativity, the behavior of the anisotropy parameter $A_p$ in a Bianchi type I geometry filled with a barotropic fluid is given by a general relation of the form $A_p(t)\propto K^2/\dot{V}^2(t)$. Hence, in any expanding Universe with $\dot{V}(t)$ a monotonically increasing function of time $\forall t\geq 0$, the anisotropy parameter will tend, in the large time limit, to zero. Thus, in standard general relativity any Bianchi type I Universe ends in isotropic phase. Anisotropic Bianchi type I universes can be given by the following physical metric
\begin{equation}
g_{\mu \nu }dx^{\mu }dx^{\nu }=-dt^{2}+e^{2\Omega }\left[ e^{2\left( \beta
_{+}+\sqrt{3}\beta _{-}\right) }dx^{2}+e^{2\left( \beta _{+}-\sqrt{3}\beta
_{-}\right) }dy^{2}+e^{-4\beta _{+}}dz^{2}\right]
\end{equation}
and auxiliary metric
\begin{equation}
q_{\mu \nu }dx^{\mu }dx^{\nu }=-X^{2}dt^{2}+Y^{2}\left[ e^{2\left( \bar{%
\beta}_{+}+\sqrt{3}\bar{\beta}_{-}\right) }dx^{2}+e^{2\left( \bar{\beta}_{+}-%
\sqrt{3}\bar{\beta}_{-}\right) }dy^{2}+e^{-4\bar{\beta}_{+}}dz^{2}\right]
\end{equation}
with $\Omega$, $X$, $Y$, $\beta _{\pm}$
and $\bar{\beta} _{\pm}$ functions of $t$ only. As a measure of anisotropy the quantity $I=d\beta /d\Omega =\sqrt{\dot{\beta}_{+}^{2}+\dot{\beta}_{-}^{2}}/H=c/H\left( \lambda e^{3\Omega }+\kappa \rho _{0}e^{-3w\Omega }\right) $ was used, where $c$, $\lambda $ and $\rho _{0}$ are constants. This definition of the anisotropy measure is different from the definition of the anisotropy parameter $A_p$, given by Equations~(\ref{an}) and (\ref{anpar}), used in the present paper. Due to the presence of the density-dependent quantities $A$ and $B$, the anisotropy parameter $A_p$ has a direct dependence on the energy density and pressure of the cosmological fluid, and on their initial conditions at the beginning of the cosmological expansion. In the high density regime, these initial conditions fully determine the evolution of the anisotropy, and some particular sets of initial conditions could lead to the presence of a small residual anisotropy in the present day universe. These anisotropy remnants may explain the observational data on the cosmological anisotropy in the CMB found by the Planck satellite \cite{Plancka,Planckb,Planckc}.}

The cosmological behavior of the dust Bianchi type I Universes differs fundamentally from the high density case. The initially anisotropic cosmological system ends in isotropic and homogeneous state, and this dynamics is independent of the assumed initial conditions for the density. The cosmological expansion of the dust fluid determines its transition from an initial anisotropic state to isotropic state.

In our analysis, we have considered the case of the cosmological fluids with isotropic pressure distribution $p_1=p_2=p_3=p$ only. More general models with anisotropic pressure distributions $p_1\neq p_2\neq p_3$, implying $B_1\neq B_2 \neq B_3$, can also be obtained in a similar way. By assuming that all pressure components satisfy a barotropic equation of state so that $p_i=p_i(\rho)$, the energy conservation equation in the physical $g$ metric takes the form
\be
\dot{\rho }+H^{(g)}\left(3\rho +p_1+p_2+p_3\right)=0
\ee while in the $q$-space the conservation equation becomes
\begin{equation}\label{92}
\dot{\rho}+\left( H+\sum_{i=1}^{3}\frac{\dot{B}_{i}}{B_{i}}-\frac{\dot{A}}{A}%
\right) \left( 3\rho +\sum_{i=1}^{3}p_{i}\right) =0
\end{equation}

The gravitational field equations with anisotropic pressure in the $q$ metric take the form
\begin{equation}\label{93}
3\dot{H}+\sum_{i=1}^{3}H_{i}^{2}=\frac{1}{\kappa }\left( 1-\frac{A}{%
\prod_{i=1}^{3}B_{i}}\right) ,\frac{1}{V}\frac{d}{dt}\left( VH_{i}\right) =\frac{1}{\kappa }\left( 1-\frac{1}{AB_{i}}\right) ,i=1,2,3
\end{equation}

From Equation~(\ref{93}) it follows that in the case of an anisotropic pressure distribution, $V$ satisfies \mbox{the equation}
\begin{equation}\label{94}
\frac{1}{V}\frac{d}{dt}\left( 3HV\right) =\frac{\ddot{V}}{V}=\frac{1}{\kappa
}\left( 3-\frac{1}{A}\sum_{i=1}^{3}\frac{1}{B_{i}}\right)
\end{equation}

Once the equations of state of the anisotropic pressures are known, $p_{i}=p_{i}\left( \rho \right) $, $i=1,2,3$, so that $B_{i}=B_{i}(\rho )$, $i=1,2,3$, Equations~(\ref{92}) and (\ref{93}) determine the full dynamics of the Bianchi type I cosmological model with anisotropic pressure distributions.

To conclude, in the present paper we have found that in the framework of the Eddington-inspired Born--Infeld gravity theory Bianchi type I Universes present complex dynamics, and in particular they do not always isotropize. The nature of the cosmological evolution strongly depends on the assumed initial conditions for the matter density and for the Hubble function. On the other hand, the presence of remnant anisotropy from the high density era of the Universe history may provide some clear cosmological signatures that could help in discriminating between the Eddington-inspired Born--Infeld gravity theory and standard general relativity.

\section*{\noindent Acknowledgments}\vspace{12pt}

We thank the anonymous referees for their very helpful comments and
suggestions that helped us to significantly improve our manuscript.
FSNL acknowledges financial support of the Funda\c{c}\~{a}o para a Ci\^{e}ncia e Tecnologia through an Investigador FCT Research contract, with reference IF/00859/2012, funded by FCT/MCTES (Portugal), and grants CERN/FP/123618/2011 and EXPL/FIS-AST/1608/2013.

\section*{\noindent Author Contributions}
\vspace{12pt}

{The authors have all contributed to the analytical calculations and plots of the paper.}

\section*{\noindent Conflicts of Interest}
\vspace{12pt}

{The authors declare no conflicts of interest.}

\bibliographystyle{mdpi}
\makeatletter
\renewcommand\@biblabel[1]{#1. }
\makeatother


\end{document}